# A Machine Learning Approach for DDoS Detection on IoT Devices

Alireza Seifousadati
*Department of Industrial Engineering*
*Iran University of Science and Technology*
Tehran, Iran
alirezasarvar.as@gmail.com

Saeid Ghasemshirazi
*Department of Industrial Engineering*
*Iran University of Science and Technology*
Kerman, Iran
saeidgs@yahoo.com

Mohammad Fathian
*Department of Industrial Engineering*
*Iran University of Science and Technology*
Tehran, Iran
fathian@iust.ac.ir

*Abstract*—In the current world, the Internet is being used almost everywhere. With the rise of IoT technology, which is one of the most used technologies, billions of IoT devices are interconnected over the Internet. However, DoS/DDoS attacks are the most frequent and perilous threat to this growing technology. New types of DDoS attacks are highly advanced and complicated, and it is almost impossible to detect or mitigate by the existing intrusion detection systems and traditional methods. Fortunately, Big Data, Data mining, and Machine Learning technologies make it possible to detect DDoS traffic effectively. This paper suggests a DDoS detection model based on data mining and machine learning techniques. For writing this paper, the latest available Dataset, CICDDoS2019, experimented with the most popular machine learning algorithms and specified the most correlated features with predicted classes are being used. It is discovered that AdaBoost and XGBoost were extraordinarily accurate and correctly predicted the type of network traffic with 100% accuracy. Future research can be extended by enhancing the model for multiclassification of different DDoS attack types and testing hybrid algorithms and newer datasets on this model.

## I. Introduction

Internet of Things (IoT) was a revolutionary technology and had become more and more beneficial in recent years. In today's world, IoT plays a crucial role in our lives. It is being used in almost every existing field like smart homes, smart cities, smart grids, autonomous vehicles, hospitals, manufacturing plants, etc. The main goal of IoT technology is to make human life more manageable and smarter by merging physical devices and digital intelligence[1,2,3,5]. IoT devices can gather data and share them from anywhere and at any time with the help of the Internet. These data are being proceeded and analyzed inside an integrated platform and will be accessible for other IoT devices. It is estimated that there are approx. 10.07 billion IoT devices connected via the Internet in 2021, and this number will reach 24.1 billion by 2030 [5]. Therefore, a large amount of data is being transferred between these interconnected devices, and it is essential to maintain this flow of data and protect it from cyber-attacks[6].

The security threats to IoT devices and networks can be sorted into six different categories: Denial of Service (DoS), Bogus Information, Eavesdropping, Impersonate, Hardware Tempering, and Message Suspension [3]. Among all the threats, DoS and Distributed DoS (DDoS) attacks, which are the more advanced version of Dos and they are more complicated to detect or mitigate, are the most dangerous and destructive method to take over IoT [7], [8]. In this type of attack, the attacker's purpose is to encumbrance the service by transmitting large volumes of data traffics, hence the service provider can't handle it, and legitimate users and devices will face problems with receiving services due to engendered disturbance [9]. There are different types of DDoS attacks with different characteristics and features. The most known types of DDoS attacks are TCP Flood, SYN Flood, UDP Flood, ICMP Flood, HTTP Flood, Ping of Death, NTP Amplification, DNS Flood, and Zero-Day DDoS [10].

As the usage of IoT technology is growing every day, it is becoming an inseparable part of our lives. However, cybersecurity threats, especially DDoS attacks, are major concerning subjects and obstacles on their way, as mentioned above [27]. Therefore, DDoS detection attracted the attention of researchers in the past few years. Many researchers proposed detection approaches for DoS/DDoS detection mostly based on machine learning algorithms [24,25,26,28].

In this research, we presented a machine learning DDoS detection model for comparing different machine learning algorithms; the main purpose of this article is to determine, which machine learning algorithm is the most accurate one, to classify the CICDDoS2019 [11], [12] dataset into binary classification within `Attack` and `Benign` classes. The architecture of our study is shown in figure 1, and the main contributions of this study are as follows:

- Providing a novel approach to feature selection and dimension reduction on the CICDDoS2019 Dataset.
- Proposing a detection model that classifies DDoS attacks more accurately and faster compared to current state-of-the-art models.
- Specifying the most proper machine learning algorithm for DoS/DDoS detection in terms of effectiveness and efficiency by examining them inside the model.
- Determining the essential features of the CICDDoS2019 Dataset that have the highest impact on DDoS prediction.

The rest of the research is listed in the following order: In section 2, the related works and other researcher's studies about machine learning techniques for DDoS detection were surveyed. In section 3, we discussed our proposed model and details of experimental environment implementation, evaluation metrics, and analyzed the results, and compared them with previous related researches results—section 4 contains the conclusion of our study. Moreover, finally, in section 5, we suggested a continuant path for our research and future works.

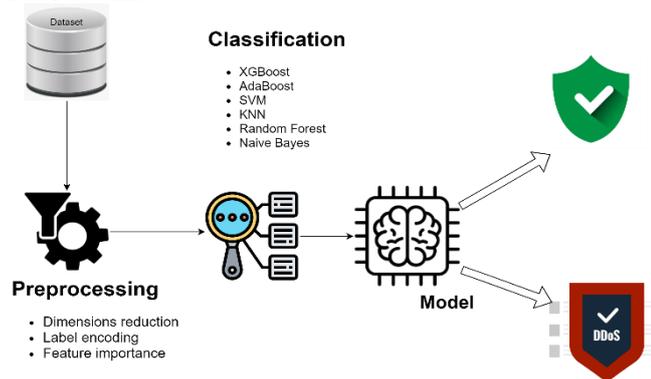

Fig. 1. The architecture of the DDoS Detection Framework

## II. BACKGROUND AND RELATED WORKS

### A. DoS and DDoS Attacks

As mentioned in the previous section, DoS attacks and their more advanced version, DDoS attacks, are among the most important security threats to the current digital world. In this type of attack, the attacker tries to overwhelm the service provider's sources by sending and receiving a large amount of network traffics, which consequently leads to service disturbance and prevents the legitimate users from utilizing their respective services. There are different types of DDoS attacks with different characteristics. Since there are different detection and mitigation methods for each type of DDoS attack, it's important to distinguish different types of attacks with their particular features. The most known kinds of DoS/DDoS attacks are explained in the following paragraph.

- **TCP Flood:** In this type of attack, the attacker exploits a part of TCP's three-way handshake to consume target resources and render it unresponsive[10].
- **SYN Flood**: In an SYN Flood attack, the attacker sends consecutive SYN packets to every target's port, using a fake IP address.
- **UDP Flood**: This is a DoS attack in which an attacker selects random target ports and then overwhelms them with IP packets containing UDP data diagrams.
- **ICMP Flood**: Internet Control Message Protocol or ICMP Flood, also known as Ping Flood, is a type of DoS attack in which the attacker attempts to make the target out of reach of normal traffic by flooding the targeted device with ICMP echo-requests.
- **HTTP Flood**: In this type of attack, the attacker overwhelms the targeted server by sending HTTP POST or GET requests.
- **Ping of Death**: In this type of attack, the attacker uses simple ping command to destabilize, crash or freeze the targeted devices by sending malformed or oversized packets.
- **NTP Amplification**: In this type of DDoS attack, the attacker exploits Network Time Protocol (NTP) servers which are publicly accessible, to overwhelm the target with UDP traffic.
- **DNS Flood**: Domain Name System or DNS Flood is a type of DDoS attack in which the attacker floods one or more particular domain's DNS servers to hamper the DNS resolution of the resource records of that domain.

### B. Machine Learning Techniques for DDoS Detection

There are various techniques for DDoS detection. However, traditional ones are becoming obsolete due to the new complicated attack types. Using data mining and machine learning techniques is the most efficient way to detect DDoS attacks and recently attracted the attention of researchers. In these kinds of techniques, a dataset is gathered from a simulation or real attack environment; then, the researchers try to extract the operative features from raw data. Subsequently, the researchers use the machine learning algorithms to train a detection model, and after that, they evaluate the performance of their model to determine whether their detection model is qualified for DDoS detection or not. A list of common machine learning algorithms for DDoS detection is available in the following paragraph.

- **Naïve Bayes**: A supervised learning and classification algorithm based on Bayes Theorem and assumes features are independent [15, 19].
- **SVM**: Supported Vector Machine is a supervised learning model which uses a classification algorithm for two-class classification problems [13, 16, 17, 20, 22].
- **AdaBoost**: Adaptive Boosting algorithm is an ensemble learning algorithm that learns from weak algorithms' faults and tries to optimize them and also make them robust classifiers gradually. In each iteration, the classifier becomes better, which is its main advantage over random predictions [23].
- **XGBoost**: Xtreme Gradient Boosting is a classification algorithm that tries to optimize the model's accuracy by minimizing the error in each iteration [15, 23].
- **KNN**: K-Nearest Neighbors is a supervised classification algorithm that uses k-closest training instances as input [17, 20, 22].
- **Random Forest**: It is an ensemble learning method for classification, and it consists of a set of decision trees that are randomly selected for training; and in the end, the final vote will be the consequence of all these trees [14, 15, 19, 20, 23].

### C. Related Works

Chuyu She et al. [13] proposed a model to distinguish normal users from botnets which are used to perform DDoS attacks on the application layer based on seven extracted features from user sessions. They used a one-class SVM algorithm on their gathered Dataset and concluded their model was effective application layer DDoS detection.

Shrikhand Wankhede et al. [14] proposed a model to detect DoS attacks based on machine learning and neural networks, then tried to maximize their model's accuracy compared to similar detection models by setting the optimum value of parameters. They achieved an accuracy of 99.95% via Random Forest algorithm with 500 trees and 50% training dataset on CIC IDS 2017 dataset.

Kian Son Hoon et al. [15] examined and analyzed different supervised and unsupervised machine learning algorithms which were being used for DDoS detection by previous researchers. They have also found the optimum value of hyperparameters that could maximize the accuracy of algorithms. The main contribution of their research is the introduction of a parameter called P(A), which is used as a threshold for better decision-making as to the training time. They have tested different algorithms on NSL-KDD Dataset and found out that the supervised algorithms like Random Forest, Gradient Boost, and Naïve Bayes have better performance in terms of accuracy and training time.

Çaḡatay Ates et al. [16] proposed a DDoS detection system based on request packet header relations. They performed experiments on real extracted data and the Caida dataset and used Entropy and Modularity concepts and the SVM algorithm. They found out that the higher accuracy is achieved by using the Entropy concept in UDP connections and Modularity concepts in TCP connections.

Shi Dong and Mudar Sarem [17] also proposed two new algorithms, DDAML and DDADA, based on KNN and the degree of attack concept. They gathered their Dataset from a simulation environment and generated DDoS traffic with hping3, and tested their proposed algorithms as well as other traditional machine learning algorithms like SVM, KNN, and Naïve Bayes. After comparing the results of ROC curves, they found out that their proposed algorithms have better performance than the existing ones.

S.Sumathi and N.Karthikeyan [18] compared different traditional and hybrid machine learning algorithms. They have tested these algorithms on KDDcup99 and DARPF datasets and found that Decision Trees and Fuzzy C-Means perform better than the others. Fuzzy C-Mean algorithm could detect DDoS traffic with an accuracy of 98.7% and with a detection time of 0.15 seconds.

Ajeetha G and Madhu Pryia G [19] developed a DDoS detection system based on machine learning techniques and traffic flow traces. They have tested Naïve Bayes and Random Forest algorithms on gathered datasets from Sans and Isna and discovered that the Naïve Bayes algorithm with an accuracy of 90.90% is more accurate than the Random Forest algorithm with 78.18% accuracy.

Khadijeh Wehbi et al. [20] reviewed the related studies on DDoS detection in the IoT environment and then proposed three new approaches using SVM, KNN, LPA, and QDA algorithms and tested these approaches on CAIDA, 1999 DARPA Intrusion Detection Dataset, and their simulated environment. Their contribution was a novel classification for feature extraction and proposed a seven-layer sequential model for DDoS detection. They have also introduced two new criteria for preventing the wrong detection of normal traffic as DDoS traffic, which is a common phenomenon for machine learning-based DDoS detection. Finally, they discovered that all three proposed approaches recorded acceptable performance, and Random Forest was the most accurate algorithm with an accuracy of 99.99%.

Esra SÖĞÜT et al. [21] proposed a DDoS detection system based on data mining techniques and machine learning algorithms. They tested different machine learning algorithms on this system and KDDCUP99 and compared them in terms of speed and accuracy. They empirically found the optimum value of some hyperparameters like 10 for Cross-Validation Ratio and 66% of Dataset for training model size. Based on this research, the J48 algorithm has the highest success rate of correct DDoS attacks detection.

Wan Nur Hidayah Ibrahim et al. [22] proposed a multilayered framework using machine learning algorithms to detect Botnets that are being used to perform DDoS attacks. They used a new approach for feature extraction, classification, and hyperparameter setting and tested KNN, SVM, and MLP algorithms on CTU-13 Dataset with their proposed framework. They discovered that, unlike previous researchers' suggestions, Oversampling technique could not improve the accuracy of algorithms. KNN algorithm recorded the highest accuracy of 91.51% inside their proposed framework.

Tarun Dhamor et al. [23] worked on DDoS detection on IoT devices. First, they used a new approach for data preprocessing on the CICDDoS2019 Dataset. Then they evaluated the performance of different machine learning algorithms for detecting DDoS traffic on their preprocessed Dataset. They ultimately discovered that machine learning techniques are effective for detecting DDoS attacks, and Random Forest, with an accuracy of 99.24%, was the most accurate algorithm among the tested algorithms.

III. METHODOLOGY

*A. Proposed Study*

First of all, we reviewed the related works of other researchers to determine the most common machine learning algorithms used for detecting Dos/DDoS attacks. Then we proposed a model to compare these algorithms in terms of effectiveness and speed. Next, we used the latest available Dataset, CICDDoS2019, as an input. After preprocessing the data, we tested the most popular machine learning algorithms and captured the data results. Finally, we specified the most important features of the CICDDoS2019 Dataset that have the highest impact on DDoS prediction for the first time.

*B. Dataset*

In this study, we used the CICDDoS2019 Dataset, which is the latest available Dataset in the context of DDoS attacks and has improved most of the shortcomings of the previous Dataset. This Dataset contains both Reflection-based and Exploitation-based DDoS attacks using TCP/UDP-based protocols at the application layer. The main benefit of using this Dataset is that it has proposed a new taxonomy, including new attack types. As a result, there are different categories of DDoS attack types which are labeled as 'PortMap,' 'NetBIOS,' 'LDAP,' 'MSSQL,' 'UDP,' 'UDP-Lag,' 'NTP,' 'DNS,' 'SNMP,' 'SSDP,' 'WebDDoS' and 'TFTP' and normal traffic which is labeled as 'BENIGN.' Network traffic data with their respective labels and traffic features which are extracted by CICFlowMeter-V3, are saved in a CSV file and available for free.

*C. Data Preprocessing*

The CICDDoS2019 Dataset is about 3GB and needs a high amount of processing resource for testing in our model; thus, it could not be directly used in our study and had to be reduced while maintaining important features and a sufficient amount of records. For data preprocessing, we used the Google Colab environment, which is developed by Google and allows everyone to write and execute python code via browser.

Therefore, it is a highly valuable tool for machine learning and data analysis researchers. We also used Pandas, Scikit-Learn, and Numpy libraries in our preprocessing work. The operation steps for data preprocessing can be summarized as below:

1. Replacing null and infinite values: The Dataset contains many infinite and null values. Therefore, we replaced null values with an average value and also infinite values with maximum values.

2. Removing remaining null values: This Dataset also contains non-numerical type null values, which cannot be replaced, and since there are enough records, as a consequence, these null values can be removed too.

3. Encoding categorical columns: Encoding non-numerical values to numerical values is essential for preparing the Dataset for experimental operations.

4. Removing columns with zero variance: When a column's variance or standard deviation is equal to zero, it means that all the values are the same, so that feature has no impact on the result.

5. Removing columns with low correlation: For dimension reduction, overfitting prevention, and making our model faster, it is necessary to remove negligible features.

*D. Machine Learning Algorithms*

We reviewed related works to determine machine learning algorithms used for DDoS attack detection by other researchers. Naïve Bayes, SVM, KNN, Random Forest, XGBoost, and AdaBoost were the most used algorithms in the literature, and these algorithms submitted great performance in DDoS detection experiments. We used these algorithms in our experimental model and then evaluated them to specify the best ones. In the following sections, the evaluation metrics and the results will be explained.

*E. Evaluation Metrics*

For comparing experimented algorithms, we used Accuracy Score, F1-Score, ROC Curve, and Training Time, and for specifying the most important features, we used Feature Importance.

- Accuracy Score measures the ratio of true predicted labels to a total number of labels. Since our Dataset might be unbalanced, just using Accuracy Score is not a good choice. The formula of Accuracy Score is mentioned below:

$$Accuracy = \frac{(TP+TN)}{(TP+FP+TN+FN)} \quad (1)$$

- F1-Score measures the harmonic average of Precision and Recall. It is a suitable evaluating metric to be accompanied by the Accuracy Score as it considers False Positive and False Negative. Formulas of Precision, Recall, and F1-Score, are mentioned in the following:

$$Precision = \frac{TP}{(TP+FP)} \quad (2)$$

$$Recall = \frac{TP}{(TP+FN)} \quad (3)$$

$$F1 - Score = \frac{2*Precision*Recall}{(Precision+Recall)} \quad (4)$$

- ROC Curve or Receiver Operating Characteristic Curve evaluates the model performance by considering False Positive and False Negative Rates.
- Training Time is a metric to determine the speed and agility of the model.
- Feature Importance calculates a correlation between each feature with the predicted label.

*F. Results And Analysis*

This section shows the result of the comparison between selected algorithms on our experimental model and CICDDoS2019 Dataset and analyzes the results. As you can see in table I and figure 2, XGBoost, and AdaBoost are the most accurate algorithms with an accuracy of 100% and an F1-Score of 1. XGBoost also recorded a slightly better Training Time than AdaBoost. Random Forest, KNN, and SVM also had acceptable accuracy of 99.94%, 99.94%, and 99.35%, and F1-Scores of 0.9942, 0.9936, and 0.9306, respectively. In our experiment, unlike previous researches, Naïve Bayes recorded an F1-Score of 0.7098, despite an Accuracy of 98.21%, which is not acceptable. Therefore, it doesn't seem to be a good algorithm for DDoS detection.

Furthermore, Naïve Bayes has a high False Positive Rate, which means that this algorithm classifies BENIGN traffic as ATTACK traffic wrongly. The reason that Naïve Bayes recorded a high Accuracy Score is that the Dataset is unbalanced and the number of Attack records is extremely more than the number of Benign records; thus, the number of False Positive classifications are not shown up in this metric, but we can understand this from F1-Score and Recall metrics. As we have seen, all selected machine learning algorithms except Naïve Bayes performed well in terms of efficiency and effectiveness. The main reason that Naïve Bayes didn't have acceptable performance was that this algorithm is based on Bayes Theorem, which assumes the feature as being independent and, in our Dataset, features were not wholly independent. Fig. 2 shows the top 10 Important Features that have the highest impact on predicting the class of network, respectively.

Table I. Detection Evaluation Results

| Algorithm | Evaluation Method | | |
|---|---|---|---|
| | *Accuracy* | *F1-Score* | *Training Time* |
| Naïve Bayes | 98.21% | 0.7098 | <1s |
| SVM | 99.35% | 0.9306 | 109s |
| XGBoost | 100% | 1.00 | 11s |
| AdaBoost | 100% | 1.00 | 14s |
| KNN | 99.94% | 0.9936 | 22s |
| Random Forest | 99.94% | 0.9942 | 4s |

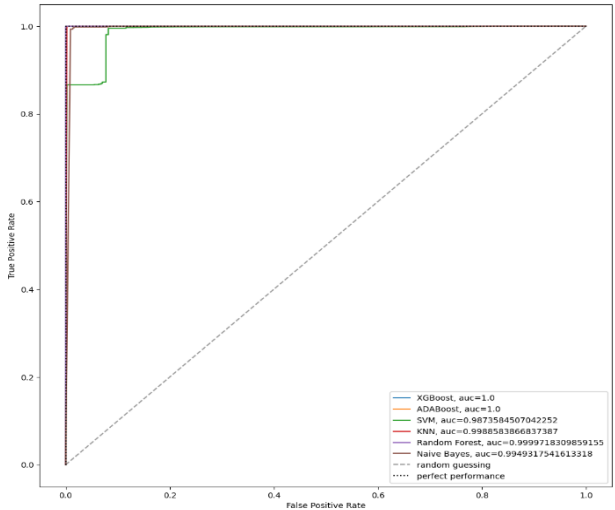

Fig. 2. ROC Curves to Evaluate Binary Classification Algorithms

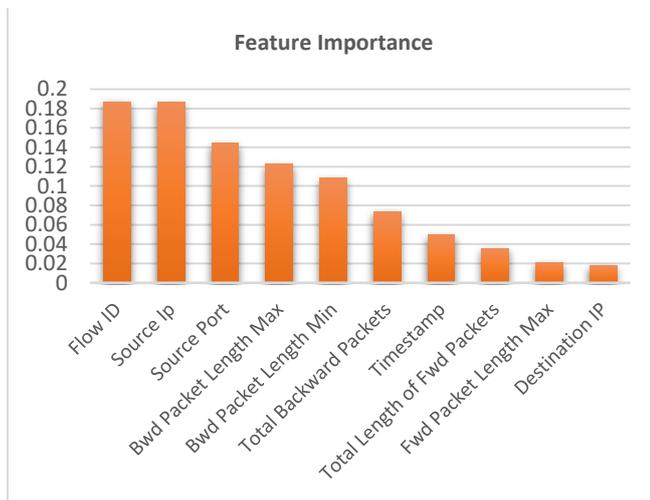

Fig. 3. Top 10 Feature Used for DDoS Detection

## IV. CONCLUSION

This paper has proposed a DDoS detection model and implemented the most popular machine learning algorithms such as Naïve Bayes, SVM, AdaBoost, XGBoost, KNN, and Random Forest, for binary classification of CICDDoS2019 network traffic into `Benign` and `Attack` classes. All of the experimented algorithms, except the Naïve Bayes algorithm, efficiently classified network traffic to Benign and Attack classes. AdaBoost and XGBoost were extremely accurate, with an Accuracy Score of 100% and F1-Score of 1. XGBoost also provides slightly better training and detection time than AdaBoost.

This study also specifies the top 10 most important features for DDoS detection, which have the highest impact on successful prediction. This is a substantial work since selecting the most pivotal features and removing non-significant ones would help the detection model to be trained better and have higher accuracy and speed and also would prevent the overfitting of the model.

## V. FUTURE WORKS

This study is focused on proposing a detection model for classifying Benign and DDoS attacks traffics on the CICDDoS2019 Dataset. In our future works, we will enhance this model so that we would be able to multi-classify the attack types. We will also test other algorithms and hybrid approaches to maximize the efficiency and effectiveness of our model. Testing this model on newer datasets could be considered as our other upcoming research.